\documentclass[prl,aps,reprint,amsmath,floatfix,superscriptaddress]{revtex4-1}
\usepackage{graphicx}
\usepackage{bm}
\usepackage{dcolumn}
\newcolumntype{d}[1]{D{.}{.}{#1}}

\newcommand{\br}{\bm{r}}
\newcommand{\bR}{\bm{R}}
\newcommand{\bP}{\bm{P}}
\newcommand{\bX}{\bm{X}}

\begin{document}

\title{
  Pair potential with submillikelvin uncertainties and nonadiabatic treatment
  of the halo state of helium dimer}
	 
\author{Micha\l\ Przybytek}
\email{mitek@tiger.chem.uw.edu.pl}
\affiliation{Department of Chemistry, University of Warsaw, Pasteura 1,
  02-093 Warsaw, Poland}

\author{Wojciech Cencek}
\affiliation{Department of Physics and Astronomy, University of Delaware, 
  Newark, DE 19716, USA}


\author{Bogumi\l\ Jeziorski}
\affiliation{Department of Chemistry, University of Warsaw, Pasteura 1,
  02-093 Warsaw, Poland}



\author{Krzysztof Szalewicz}
\affiliation{Department of Physics and Astronomy, University of Delaware, 
  Newark, DE 19716, USA}

\date{\today}                 

\begin{abstract}
  The pair potential for helium has been computed with accuracy improved by  
  an order of magnitude relative to the best previous determination.
  For the well region, its uncertainties are now below 1~millikelvin.
  The main improvement is due to the use of explicitly correlated wave functions
  at the nonrelativistic Born-Oppenheimer (BO) level of theory.
  The diagonal BO and the relativistic corrections were obtained
  from large full configuration interaction calculations. The nonadiabatic
  perturbation theory was used to predict the properties of the halo state 
  of helium dimer. Its binding energy and the average value of interatomic
  distance are found to be 138.9(5)~neV and 47.13(8)~\AA.
  The binding energy agrees with its first experimental determination
  of 151.9(13.3)~neV [Zeller et al., PNAS {\bf 113}, 14651 (2016)]. 
\end{abstract}

\maketitle

\newpage 


Helium is expected to become an important medium in determining thermodynamic
metrology standards and the future system of SI units
\cite{Fischer:16,Moldover:16}. 
Several elements of such standards will be established by \emph{ab initio}
quantum mechanical calculations
\cite{Schmidt:07,Egan:15,Cencek:12,Piszczatowski:15,Gavioso:16}. 
An important theory input is the helium pair potential. Its knowledge is
required to account for the imperfection of helium gas and the necessary
extrapolations to zero pressure \cite{Moldover:16}. 
The more accurate this potential is, the 
smaller will be the uncertainties of the resulting standards.

There are other reasons of interest in the helium pair potential.
The dimer composed of $^4$He atoms, $^4$He$_2$, has a single very weakly bound
vibrational state---an example of a quantum halo state---where atoms move  
mainly in  the classically forbidden  tunneling region of the configuration
space \cite{Riisager:94}. This state was the subject of several experimental
investigations
\cite{Luo:93,Luo:96,Schollkopf:94,Schollkopf:96,Grisenti:00,Zeller:16}.
We present here the development of a new potential with uncertainties
reduced by   an order of magnitude compared to the previous
most accurate determination \cite{Przybytek:10}. This potential and the 
nonadiabatic perturbation theory \cite{Pachucki:08a}, accounting for 
the coupling of the electronic and nuclear motion, are used to obtain
an accurate theoretical prediction of the properties of the halo state.

The potential of Ref.~\cite{Przybytek:10} contained
the Born-Oppenheimer (BO) component from  Ref.~\cite{Jeziorska:07}.
Its uncertainty, amounting to several
millikelvin (mK) in the well region, was  due to
the slow convergence of a part of the wave function expanded in terms
of orbital products.
Since it is impossible to converge the orbital expansion
sufficiently well \cite{Cencek:17},
we now follow Refs.~\cite{Cencek:93,Komasa:96}
and expand the BO wave function
using the four-electron explicitly correlated Gaussian (ECG) basis.
Several improvements to the approach of Refs.~\cite{Cencek:93,Komasa:96}
that have been made recently \cite{Cencek:05,Cencek:08,Patkowski:08}
enabled us to perform highly accurate ECG calculations
for 46 values of the interatomic distance $R$.

In Ref.~\cite{Przybytek:10}, the BO potential of Ref.~\cite{Jeziorska:07}
was combined with the adiabatic (diagonal BO), relativistic,
and quantum electrodynamics (QED) contributions, as well as
with an appropriate retardation correction \cite{Przybytek:12}.
Its uncertainties were almost entirely determined by the uncertainties of
the BO component. 
With the much improved BO potential computed in the present work, the accuracy 
of the adiabatic and relativistic components from Ref.~\cite{Przybytek:10}
became insufficient. Therefore, we decided to recompute these components using 
different methodologies, providing higher accuracy and better error control. 
  
Recently, the wave function of $^4$He$_2$ has been measured via the Coulomb 
explosion technique \cite{Zeller:16}, which enabled the first experimental 
determination of its very small binding energy (151.9$\pm$13.3~neV).
The most precise calculation for this state was performed \cite{Przybytek:10}
in the adiabatic approximation giving the binding energy
$D_0$=136.6$\pm$2.9~neV when nuclear masses are used to solve
the vibrational problem (as required by the mathematical derivation of
the adiabatic approximation) or 139.2$\pm$2.9~neV when
the atomic masses are used (as suggested by physical intuition).
The average interatomic separation $\langle R\rangle$ obtained
with these  masses were 47.50$\pm$0.46~\AA\ and 47.09$\pm$0.46~\AA,
respectively, in a minor disagreement with the experimental value of
52$\pm$4~\AA\ \cite{Grisenti:00}. 
To resolve this ambiguity, in the present work we have used the nonadiabatic 
perturbation theory \cite{Pachucki:08a} to account for the coupling of
the electronic and nuclear motion. This requires the calculation of
an effective $R$-dependent vibrational mass and of a nonadiabatic correction
to the potential \cite{Pachucki:08a}. 
We have developed methods to compute these quantities for many-electron
diatomics and report the results in this communication.
To our knowledge, such nonadiabatic calculations have not been performed
earlier for systems with more than two electrons.
   

The ECG wave function  employed by us has the form
\begin{equation}
  \label{Psi}
  \Psi = \mathcal{A} \, \Xi (1+\hat{\imath}) \bigg[
    c_0 \phi_0 +
    \sum_{k=1}^K \, c_k \, \phi_k (\br_1,\br_2,\br_3,\br_4) \bigg] ,
\end{equation}
where $\mathcal{A}$ is the antisymmetrizer,
$\Xi$ is the product of two-electron singlet spin functions,
$\hat{\imath}$ is the inversion through the center of He$_2$, and $\phi_k$,
$k${}$>$0, are the ECG basis functions:
\begin{equation} 
  \label{phi}
  \phi_k (\br_1,\br_2,\br_3,\br_4) =
  \prod_{i=1}^{4} e^{ -\alpha_{ki} |{\br}_i - {\bX}_{ki}|^2 } 
  \!\!\prod_{i>j=1}^{4} \!\! e^{ -\beta_{kij} |{\br}_i - {\br}_j|^2 }.
\end{equation}
The linear parameters $c_k$ and the nonlinear ones
$\alpha_{ki}$, $\beta_{kij}$, and $\bX_{ki}$=$(0,0,X_{ki})$ are
optimized
by minimizing the expectation value of the electronic Hamiltonian
$\hat{H}_{\rm el}$.
The term $c_0\phi_0$ is included to approximate
the product of spinless helium atom wave functions.
When performing the nonlinear optimization of $\Psi$,
we used the following fixed form of $\phi_0$
\begin{equation}
  \label{phi0}
 \phi_0 = {\cal S} \sum_{l=1}^{L^{\prime}} b_{l}\,
  \phi_{\alpha_l\beta_l\gamma_l}^{A} (\br_1,\br_2) \,
  \phi_{\alpha'_l\beta'_l\gamma'_l}^{B} (\br_3,\br_4) ,
\end{equation}
where ${\cal S}$=$(1+\hat{\imath}P_{13} P_{24})(1+P_{12})(1+P_{34})$,  with 
 $P_{ij}$ permuting the coordinates of the $i$th and $j$th electron,  and  
 \begin{equation}  
  \label{phi12}
  \phi_{\alpha_l\beta_l\gamma_l}^{X}\!(\br_1,\br_2) = 
  e^{ -\alpha_{l} |{\br}_1 - {\bX}|^2 }  
  e^{ -\beta_{l} |{\br}_2 - {\bX}|^2 }
  e^{ -\gamma_l |{\br}_1 - {\br}_2|^2 } , 
\end{equation}
with $\bX$=(0,0,0) for $X$=$A$ and $\bX$=(0,0,$R$) for $X$=$B$. 
The  parameters of $\phi_0$ were  
optimized by minimizing the expectation value of the sum $\hat{H}_A+\hat{H}_B$
of the atomic Hamiltonians \cite{Patkowski:08}. We have set $L'$=6788, 
obtaining the energy of two noninteracting helium atoms within
0.16~mK of the exact value of Ref.~\cite{Nakashima:08}.
After all nonlinear parameters in $\phi_k$, $k${}$>$0, were optimized
the final energy was computed with $\phi_0$ represented 
by the product of helium wave functions expanded in terms of   337 symmetrized ECG's 
of the form of Eq.~(\ref{phi12}).
The energy of two helium atoms computed with this form of  $\phi_0$
differs from the exact one by 0.02~mK.
 
The calculations were first performed for the same 16 internuclear distances
as in Ref.~\cite{Jeziorska:07}, ranging from 1 to 9~bohr. 
For each distance, $K$=2400, 3394, 4800, and 6788 term expansions of the form
of Eq.~(\ref{Psi}) were optimized.
Attempts to fit analytic functions to the computed interaction energies
have shown that the assumed grid density is insufficient
to obtain a fit to within the new, decreased uncertainty.
Therefore, we performed calculations at additional 30 values of $R$
located at 0.33 and 0.67 of the distances between the existing 16
points (with $R$ rounded to 0.01 bohr).
The nonlinear parameters for the additional values of $R$ were obtained
from the wave function of the nearest $R$ from the original set, employing
the scaling procedure proposed in Ref.~\cite{Cencek:97}.

The interaction energy, $E(K)$,  was obtained by subtracting the exact atomic 
energies \cite{Nakashima:08} from the calculated dimer energy,
so  $E(K)$ is a rigorous variational upper bound. 
To extrapolate to the complete basis set (CBS) limit,
we employed an empirical observation that the ratio 
$\eta_K ${}$=${}$\Delta_{K/\sqrt{2}}/\Delta_K  $, with 
$\Delta_K${}$=${}$E(K)${}$-${}$E(K/\sqrt{2})$,
is approximately independent of $K$. Disregarding a few outliers,
we found that the values of $\eta_K$ are between 1.32 and 3.  
We have chosen $\eta$=1.32, to determine the extrapolated interaction energy 
$E_\text{extrp}=E(6788)+\Delta_{6788}/(\eta-1)$. 
This choice, resulting in the largest magnitude of the CBS correction,
compensates for the incompleteness of the minimization for $K$=6788.
The difference of energies extrapolated with $\eta$=3 and $\eta$=1.32 
was taken as the uncertainty of $E_\text{extrp}$. 
 
The CBS-extrapolated values of the BO interaction energies
and their uncertainties are listed in Table~\ref{table:v}
for a subset of distances. 
The data for other distances are given in Supplementary Information (SI) \cite{SI}.
The BO energies reported in Ref.~\cite{Cencek:12} are
presented for comparison. At all 16 distances where energies from both
sets are available, the uncertainties overlap, so both sets
of results are consistent.  However, the present uncertainties  are
tighter by about an order of magnitude
(from 8 to 23 times for $R${}$<$7~bohr
and from 2.5 to 6 times for other distances), except at 5.6~bohr.

\begingroup
\squeezetable
\begin{table*}
  \caption{\label{table:v}
    Components of the $^4$He dimer potential in kelvin
    (1~hartree=315775.13~K) with $R$ in bohr (1~bohr=0.529177~\AA)
    and their sum 
    $V=V_\text{BO}+V_\text{ad}+V_\text{rel}+V_\text{QED}$.
     Results for other values of $R$ and the components of $V_\text{rel}$
    are listed in the SI \cite{SI}.
  }
  \begin{ruledtabular}
    \begin{tabular}{d{1}d{4.9}d{4.9}d{2.11}d{2.11}d{2.10}d{4.9}d{4.8}d{1.5}}
      \multicolumn{1}{c}{$R$}&
      \multicolumn{1}{c}{$V_\text{BO}$} &
      \multicolumn{1}{c}{$V_\text{BO}$, Ref.~\cite{Cencek:12}} &
      \multicolumn{1}{c}{$V_\text{ad}$} &
      \multicolumn{1}{c}{$V_\text{rel}$} &
      \multicolumn{1}{c}{$V_\text{QED}$} &
      \multicolumn{1}{c}{$V$} &
      \multicolumn{1}{c}{$V$, Ref.~\cite{Cencek:12}} &
      \multicolumn{1}{c}{$V_\text{ret}$} \\
      \hline
      3.0 &    3767.7341(38)    &    3767.681(71)     &       1.3847(15)    &      -0.2125(17)    &       0.09376(22)   &    3769.000(4)      &    3768.94(7)       & 0.00045 \\
      4.0 &     292.58201(86)   &     292.570(15)     &       0.10585(17)   &       0.03322(21)   &       0.00891(5)    &     292.7300(9)     &     292.719(15)     & 0.00025 \\
      5.0 &      -0.47114(36)   &      -0.4754(65)    &      -0.006992(10)  &       0.024012(25)  &      -0.00106(3)    &      -0.4552(4)     &      -0.460(7)      & 0.00015 \\
      5.6 &     -11.00072(20)   &     -11.0006(2)     &      -0.008905(10)  &       0.015403(15)  &      -0.001351(23)  &     -10.99557(20)   &     -10.9955(5)     & 0.00012 \\
      6.0 &      -9.68079(16)   &      -9.6819(23)    &      -0.007170(4)   &       0.011438(11)  &      -0.00120(4)    &      -9.67772(16)   &      -9.6788(23)    & 0.00010 \\
      7.0 &      -4.62260(10)   &      -4.6225(6)     &      -0.0033168(24) &       0.005768(4)   &      -0.00074(3)    &      -4.62089(11)   &      -4.6208(6)     & 0.00007 \\
      9.0 &      -0.98971(6)    &      -0.98984(15)   &      -0.0007328(8)  &       0.0019306(6)  &      -0.000316(29)  &      -0.98883(7)    &      -0.9890(2)     & 0.00004 \\
      12.0 &           &      -0.16592(2)   &      -0.0001261(1)  &       0.0005768(1)  &      -0.000133(26)  &      -0.16560(3)\footnotemark[1]    &      -0.16560(3)     & 0.00002 \\
    \end{tabular}
  \end{ruledtabular}
  \footnotetext[1]{Computed with the same value of $V_\text{BO}$
    as in Ref.~\cite{Cencek:12} (given in the third column).}
\end{table*}
\endgroup


In Ref.~\cite{Cencek:12}, the adiabatic correction $E_\text{ad}(R)$
was computed via numerical differentiation of the electronic wave function
with respect to nuclear positions. In our work,
we employed the method proposed   by Pachucki and Komasa
\cite{Pachucki:08a}. In a space-fixed reference frame, $E_\text{ad}(R)$ is
expressed as \cite{Pachucki:08a}
\begin{equation}
  \label{ad:dimer}
  E_\text{ad}(R) =
  \frac{\hbar^2}{m_n}\langle \nabla_{\bR}\Psi | \nabla_{\bR}\Psi \rangle 
  +\frac{1}{4m_n}\langle\Psi |\bP^2| \Psi\rangle,
\end{equation}
where $\bR$ is the vector joining the nuclei, $m_n$ is the nuclear mass,
and $\bP$ is the total electronic momentum operator.
To avoid the cumbersome differentiation of $\Psi$ with respect to $\bR$,
we obtained $\nabla_{\bR}\Psi$ by solving the equation \cite{Pachucki:08a} 
\begin{equation}
  \label{ad:diff}
  ( \hat{ H}_\text{el} - E_\text{BO}) \nabla_{\bR}\Psi =
  - (\nabla_{\bR}\hat H_\text{el} ) \Psi.
\end{equation}
The adiabatic correction to the potential is defined as 
$V_\text{ad}(R)=E_\text{ad}(R)-2E^A_\text{ad}$,
where $E^A_\text{ad} $ is the atomic adiabatic correction \cite{Przybytek:17}.
When $E_\text{ad}(R)$ and $ E^A_\text{ad}$ are computed with 
the same basis, $V_\text{ad}(R)$ vanishes at large $R$
in accord with its known asymptotic expansion \cite{Przybytek:12a}.
 
The solution $\nabla_{\bR}\Psi$ of Eq.~(\ref{ad:diff}) was obtained
by representing $\nabla_{\bR}\Psi$ and $\Psi$ as full configuration interaction
(FCI) expansions and solving linear equations for the CI coefficients.
By comparing with accurate ECG results, available at 
small $R$ \cite{Cencek:12}, we found that the orbital basis sets
d$X$Z from Ref.~\cite{Cencek:12} lead to fast convergence 
provided that they are augmented by one set of $p$ functions obtained
by taking the nuclear gradient of the contracted, 19-term $s$ orbital
already present in all d$X$Z bases of Ref.~\cite{Cencek:12}.
The d$X$Z bases augmented in this way will be referred to as the d$X$Zcp bases. 
 
$V_\text{ad}(R)$ was calculated using the d$X$Zcp bases up to
$X$=6 for 55 values of $R$, the same 46 values as in the case of
the BO potential and, additionally, for 9 larger distances.
The largest FCI calculations employed the wave functions with
$\sim\!4\times10^8$ determinants (at $D_{2h}$ symmetry).
All necessary integrals and Hartree-Fock
orbitals were computed using the \textsc{Dalton 2.0} package \cite{dalton},
while the adiabatic corrections were obtained using an FCI code  written
for the purpose of this work. The values of $V_\text{ad}(R)$ 
were extrapolated to the CBS limit assuming the $X^{-3}$ decay of the error. 
As our recommended values of $V_\text{ad}(R)$, we took the CBS limit based on
the d5Zcp and d6Zcp results with uncertainties estimated
as the absolute values of the difference between the extrapolated and the d6Zcp
result. Combining the new numerical approach and the increased size of  
basis sets (in Ref.~\cite{Cencek:12},   bases  up to  $X${}$=${}$4$   were 
  used), we reduced the   uncertainty of the adiabatic
corrections by an order of magnitude.


The relativistic component, $V_\text{rel}(R)$, 
of the potential $V(R)$ was computed for 55 values of $R$ using
the same method as in Ref.~\cite{Cencek:12}, except that we employed
basis sets with larger cardinal numbers $X$ and added $p$ functions
to improve the wave function in the vicinity of nuclei. 
Specifically, we started with the \emph{modified} d$X$Z basis sets of
Ref.~\cite{Cencek:12} (containing 21 uncontracted $s$ functions)
and augmented them by $n${}$\le$5 ``tight" $p$ functions
with exponents larger than those already present in the original 
d$X$Z basis. The bases obtained in this way will be denoted as d$X$Z+$n$p.
The exponents of these ``tight" $p$ functions are given in the SI~\cite{SI}.

To calculate expectation values of the relativistic operators, we 
used a composite approach. The main contribution (over 90\%) 
was calculated at the coupled cluster CCSD(T) level of theory \cite{Coriani:04} 
using large basis sets (up to d8Z+5p) whereas the remaining  contribution 
was included applying an additive FCI correction computed
with smaller bases (up to d6Z+5p). The CCSD(T) calculations were performed using
the \textsc{Dalton 2013} package \cite{dalton2013}, whereas at the FCI level
we used a program written for this work. 
For each internuclear distance, the relativistic potentials were obtained
as the difference between the dimer and atomic expectation values,
the latter calculated with the dimer basis to remove
the basis-set superposition error.
 
To perform CBS extrapolations, we employed the convergence laws established
in Ref.~\cite{Cencek:12}, i.e., we assumed that upon increasing
the cardinal number $X$, the error of the Breit correction decays   
as $X^{-3/2}$ and the errors of the remaining corrections as $X^{-1}$. 
The fixed-$n$ extrapolation from bases d($X-1$)Z+$n$p and d$X$Z+$n$p
will be denoted as d($X-1,X$)Z+$n$p.  
We found that the effect of the increased flexibility of the new d$X$Z+$n$p
bases on the relativistic corrections is small, especially for $n${}$>$3, 
although it improves somewhat the  convergence of the extrapolations. 
As our recommended CCSD(T) component of the relativistic corrections
we took the d(7,8)Z+5p extrapolation with uncertainties estimated
as the absolute value of the difference between  
the d(7,8)Z+5p and  d(6,7)Z+5p extrapolations.  Similarly, 
at the FCI level, we used the d(5,6)Z+5p extrapolation with uncertainties
estimated as the absolute value of the difference between
the d(5,6)Z+5p and d(4,5)Z+5p extrapolations. 
To check the basis set convergence of the FCI correction,
we also carried out FCI calculations for three distances, $R$=2, 5.6,
and 12~bohr using the d7Z+2p basis set which consists of 512 functions
(and generates $\sim\!2\times10^9$ $D_{2h}$-adapted determinants).
The results of the FCI extrapolations d(6,7)Z+2p for $R$=5.6 and 12~bohr
(where the FCI corrections are most relevant), are contained within
the proposed error bars which shows that our uncertainty estimates are reliable.
 
The calculated one- and two-electron Darwin terms
together with the ECG results for the Araki-Sucher term, $V_\text{AS}(R)$,
from Ref.~\cite{Cencek:12} were  employed  to compute 
the leading  (third-order in the fine structure constant $\alpha$) QED correction, $V_\text{QED}(R)$, using the formulas from Ref.~\cite{Cencek:12}.  Using Eq.~(19)   
from Ref.~\cite{Cencek:12}, we also estimated the $\alpha^4$ QED correction and 
found that it is at least 5 times  smaller than the uncertainties of  $V(R)$. 
Therefore this correction was  neglected.

The uncertainties of the components  of $V_\text{rel}(R)$ and of $V_\text{QED}(R)$,
as well as uncertainties of all components of $V(R)$, were added in squares.
Compared to the results from Ref.~\cite{Cencek:12}, the uncertainties of
$V_\text{rel}(R)$ were reduced by a factor 1.4--17 depending on $R$.
The uncertainties of $V_\text{QED}(R)$ remain  unchanged
as they are dominated by the uncertainty of the Araki-Sucher component.
Also the retardation correction, appropriate for the potential including
the leading QED term \cite{Przybytek:12}, is the same as
in Ref.~\cite{Cencek:12}. 
As seen in Table~\ref{table:v}, the uncertainties
assigned to all calculated post-BO corrections to the interaction potential 
are comparable or smaller than the uncertainties of the BO potential.


The computed values of
$V_\text{BO}(R)$,
$V_\text{ad}(R)$,
$V_\text{rel}(R)$, and
$V_\text{QED}(R)$ were fitted to the analytic functions of the form
\begin{equation}
  \label{fitfun}
  \sum^M_{k=1} e^{-a_kR}
  \sum_{i=I_0}^{I_1} P_{ik} R^i
  - \sum_{n=N_0}^{N_1} f_n(\zeta R) \frac{C_n}{R^n},
\end{equation}
where $f_n(x)$=$1-e^{-x}(1+x+\cdots+x^n/n!)$ is the Tang-Toennies \cite{Tang:84}
damping function, $a_k$, $P_{ik}$, and $\zeta$ are adjustable parameters,
and the summation limits [$M,I_0,I_1,N_0,N_1$] are
\mbox{[3,-1,2,6,16]} for $V_\text{BO}(R)$,
\mbox{[3,0,2,6,10]} for $V_\text{ad}(R)$,
\mbox{[3,0,2,4,8]} for $V_\text{rel}(R)$, and
\mbox{[2,0,2,3,6]} for $V_\text{QED}(R)$.
The asymptotic constants
$C_8$ for $V_\text{rel}(R)$ and $C_6$ for $V_\text{QED}(R)$
are not known and were also adjusted. The remaining constants $C_n$ were fixed
and set equal to the known literature 
values \cite{Przybytek:08,Przybytek:12a,Cencek:12,Mitroy:11}.
To impose the correct behavior of $V_\text{BO}(R)$ at $R$=0 we used the theoretical value of the beryllium atom energy   $E_\text{Be}${}$=${}$-${}$14.667356498$ hartree \cite{Puchalski:13}. 
We used the inverse squares of uncertainties $\sigma(R)$ as
the weighting factors in the least-squares fitting.
The maximum and average absolute errors of the fit are
0.92$\,\sigma$ and 0.16$\,\sigma$, respectively, for the BO component.
Similarly accurate fits were obtained for the remaining components of $V(R)$.

In order to estimate the uncertainties of physical quantities calculated
with our potential, we developed functions $\sigma_X(R)$ representing
the uncertainties of the calculated components such that their exact values
can be assumed to be contained between functions $V_X(R)\pm\sigma_X(R)$,
where $V_X(R)$ is the analytic fit of a component $X$.
We found that the functions $\sigma_X(R)$
can be represented as $\sigma_X(R)=s_0e^{-a_0R}+\sum_{i=1}^{n}s_{i}e^{-a_iR^2}$
where $n$=3, except for $V_\text{rel}(R)$ when $n$=4.
The parameters and the Fortran codes for all fits can be found
in the SI \cite{SI}.


To compute the properties of the bound state of $^4$He$_2$, we used
the nonadiabatic perturbation theory \cite{Pachucki:08a} applied successfully 
to the H$_2$ molecule and its isotopologues
\cite{Pachucki:09,Pachucki:10,Komasa:11,Pachucki:11,Pachucki:15}. 
In this theory, the energies $E$ and radial wave functions $\chi(R)$
are obtained by solving the radial equation of the form
\begin{equation}
  \label{eq:radial}
  \left[-\frac{\hbar^2}{R^2}
    \frac\partial{\partial R}
    \frac{R^2}{2\mu_\parallel(R)}
    \frac\partial{\partial R}
    \!+\!
    \frac{J(J+1)\hbar^2}{2\mu\!_\perp\!(R)R^2}
    \!+\!\mathcal{Y}(R)
    \!-\!E \right] \chi(R) \!=\! 0,
\end{equation}
where $\mu_\parallel(R)$ and $\mu\!_\perp\!(R)$ are the $R$-dependent
vibrational and rotational reduced masses
\begin{equation}
  \label{eq:mass}
  \frac1{2\mu_\parallel(R)}\!=\!\frac1{m_n}\!+\!\mathcal{W}_\parallel(R),
  \qquad
  \frac1{2\mu\!_\perp\!(R)}\!=\!\frac1{m_n}\!+\!\mathcal{W}\!_\perp\!(R),
\end{equation}
and $\mathcal{Y}(R)$ is the sum of 
$V(R)$, 
$V_\text{ret}(R)$,
and a nonadiabatic correction $V_\text{na}(R)$.  
The   expressions for the functions
$\mathcal{W}_\parallel(R)$,
$\mathcal{W}\!_\perp\!(R)$,
and $V_\text{na}(R)$ are given in Ref.~\cite{Pachucki:09}.
One can show that 
$2\mu_\parallel(\infty)$=$2\mu\!_\perp\!(\infty)$=$m_n+2m_e+4m_e^2/m_n+\mathcal{O}\left(m_e^3/m_n^2\right)$,
where $m_e$ is the electron mass. 
We employed  the known $R${}$\rightarrow${}$\infty$ limits   
and computed directly the $R$-dependent parts
$\mathcal{W}^\text{int}_\parallel(R)${}$\equiv${}$\mathcal{W}_\parallel(R)${}$-${}$\mathcal{W}_\parallel (\infty)$
and
$\mathcal{W}^\text{int}_\perp\!(R)${}$\equiv${}$\mathcal{W}_\perp\!(R)${}$-${}$\mathcal{W}_\perp\!(\infty)$
of $\mathcal{W}_\parallel(R)$ and $\mathcal{W}\!_\perp\!(R)$.

The values of the functions
$\mathcal{W}_\parallel^\text{int}(R)$,
$\mathcal{W}^\text{int}_\perp\!(R)$, and 
$V^\text{int}_\text{na}(R)${}$\equiv${}$V_\text{na}(R)${}$-${}$V_\text{na}(\infty)$
were calculated at 52 points in the range 1$\leq${}$R${}$\leq$18~bohr
using a dedicated FCI code and the same d$X$Zcp orbital basis sets as
used to calculate the adiabatic correction,
see Ref.~\cite{Przybytek:17} for the   description of
the computational algorithm. We employed basis sets
with   cardinal numbers up to $X$=6 for
$\mathcal{W}_\parallel^\text{int}(R)$ and 
$\mathcal{W}_\perp^\text{int}(R)$
and up to $X$=5 for $V^\text{int}_\text{na}(R)$.
The recommended values of
$\mathcal{W}_\parallel^\text{int}(R)$,
$\mathcal{W}^\text{int}_\perp\!(R)$, and 
$V^\text{int}_\text{na}(R)$
were obtained by extrapolations from the results computed
with two largest basis sets assuming the $X^{-3}$ convergence.
The analytic representations of these functions were obtained
by fitting the recommended values with functions of the form of 
Eq.~(\ref{fitfun}) with summation limits [$M,I_0,I_1,N_0,N_1$]
equal to
\mbox{[2,0,3,8,8]} for $\mathcal{W}_\parallel^\text{int}(R)$,
\mbox{[2,0,2,8,8]} for $\mathcal{W}^\text{int}_\perp\!(R)$, and
\mbox{[3,0,2,6,8]} for $V_\text{na}^\text{int}(R)$.
 We estimate that in the well region
the obtained fits represent the exact values 
with errors smaller than 5\%. 
Equation~(\ref{eq:radial}) was solved numerically using
the Mathematica software \cite{Wolfram}.
 
\begin{figure}
  \includegraphics[width=\columnwidth]{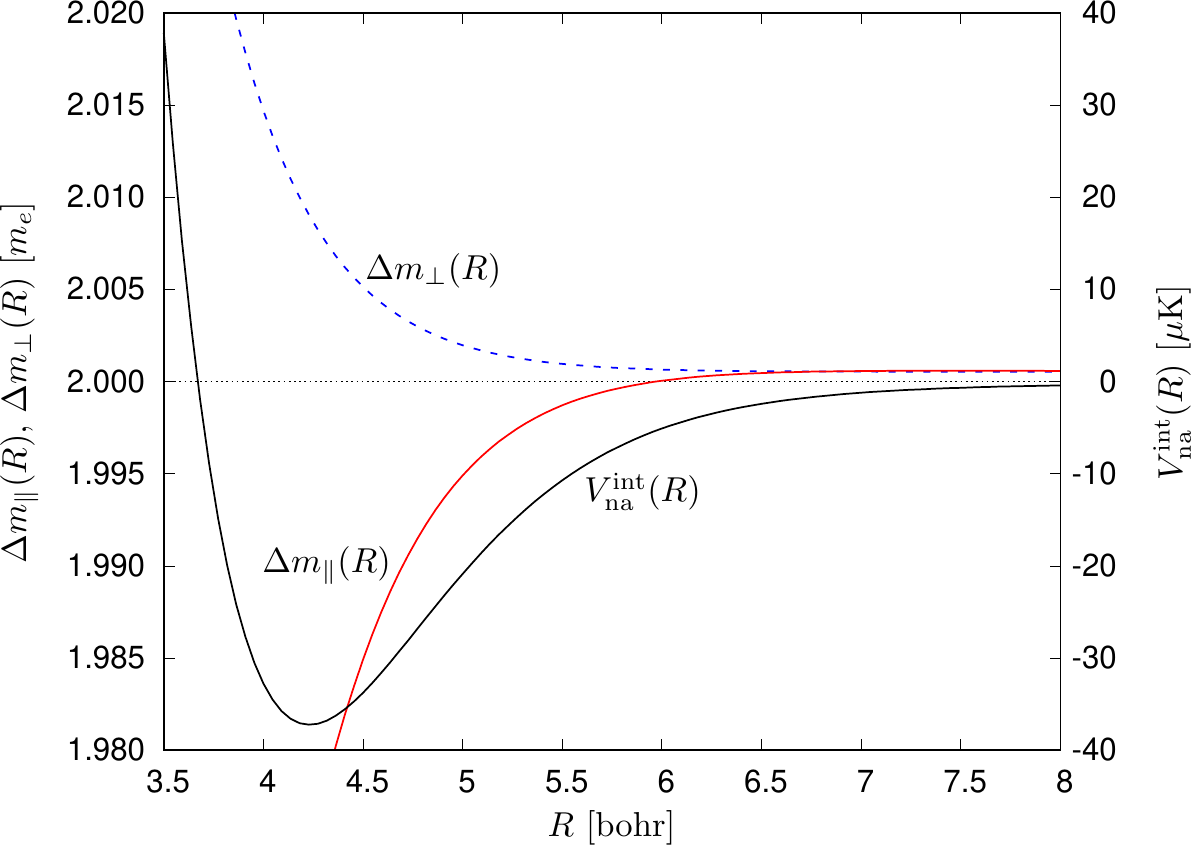}
  \caption{
    \label{figure:nad}
    The $R$-dependence of the excess masses and
    the nonadiabatic correction to the potential. 
  }
\end{figure}

The computed dissociation energy $D_0$ and the size $\langle R\rangle$
of the ($J$=0) bound state are presented in Table~\ref{table:results},
while the plots of the excess masses 
$\Delta m_\parallel(R)$=2$\mu_\parallel(R)-m_n$, 
$\Delta m\!_\perp(R)$=2$\mu\!_\perp(R)-m_n$, and of
$V^\text{int}_\text{na}(R)$
are shown in Fig.~\ref{figure:nad}.
Our results confirm earlier observation \cite{Przybytek:10}
that the adiabatic and relativistic corrections to $D_0$ and $\langle R\rangle$
are significant, but the effect of retardation is very small when
the leading relativistic and QED contributions are included in $V(R)$.
The nonadiabatic effect increases $D_0$ by 2.6~neV
and decreases $\langle R\rangle$ by 0.42~\AA,
i.e., by the same amount as does the QED  correction.
It is interesting to observe that these changes are recovered
with excellent accuracy by the adiabatic calculations with atomic masses.
We found that the difference between the nonadiabatic values of $D_0$ and
$\langle R\rangle$ and the adiabatic ones computed with atomic 
masses are only $-$0.0007~neV and 0.00011~\AA, respectively.
These differences are negligible due to the small values
$\Delta m_\parallel(R)-2m_e$ in the well region
($R${}$>$5~bohr),
as shown in Fig.~\ref{figure:nad}, but can be expected to be larger
for helium properties sensitive to the potential at smaller values of $R$.
Our results resolve the long-standing controversy
\cite{Janzen:97,Jamieson:98,Jamieson:99a}
which masses should be used in calculations for weakly bound dimers.

\begin{table}
  \caption{
    \label{table:results}
    Dissociation energy $D_0$ (in neV)
    and the average separation $\langle R\rangle$ (in \AA) for $^4$He$_2$.
    $V=V_\text{BO}+V_\text{ad}+V_\text{rel}+V_\text{QED}$.
  }
  \begin{ruledtabular}
    \begin{tabular}{ld{3.4}d{3.5}d{2.5}d{2.6}}
      &\multicolumn{2}{c}{$D_0$}&\multicolumn{2}{c}{$\langle R\rangle$} \\
      \multicolumn{1}{c}{potential} &
      \multicolumn{1}{c}{nuclear} &
      \multicolumn{1}{c}{atomic\phantom{a}} &
      \multicolumn{1}{c}{nuclear} &
      \multicolumn{1}{c}{atomic\phantom{a}} \\
      \hline
      $V_\text{BO}$                         &   145.2(5)    &   147.8(5)    &   46.20(7)    &   45.80(7)    \\
      $V_\text{BO}$+$V_\text{ad}$             &   153.5(5)    &   156.3(5)    &   45.03(7)    &   44.65(7)    \\   
      $V_\text{BO}$+$V_\text{ad}$+$V_\text{rel}$&   134.1(5)    &   136.7(5)    &   47.90(8)    &   47.48(8)    \\                
      $V$                                   &   136.7(5)    &   139.3(5)    &   47.48(8)    &   47.07(8)    \\
      $V$+$V_\text{ret}$                      &   136.3(5)    &   138.9(5)    &   47.55(8)    &   47.13(8)    \\
      $V$+$V_\text{ret}$+$\text{nonad}$        &   138.9(5)    &               &   47.13(8)    &               \\   
      $V$+$V_\text{ret}$, Ref.~\cite{Przybytek:10}          &               &   139.2(29)   &               &   47.09(46)   \\
      Exptl &
      \multicolumn{2}{c}{$151.9\pm13.3$\footnotemark[1]}&
      \multicolumn{2}{c}{$52\pm4$\footnotemark[2]} \\
    \end{tabular}
  \end{ruledtabular}
  \footnotetext[1]{Ref.~\cite{Zeller:16}}
  \footnotetext[2]{Ref.~\cite{Grisenti:00}}
\end{table}

The recommended values of $D_0$=138.9(5)~neV and
$\langle R\rangle$=47.13(8)~\AA\ agree with
the former best theoretical determinations \cite{Przybytek:10},
but have six times smaller uncertainties.
The small disagreement with the best measured value of
$\langle R\rangle$ \cite{Grisenti:00} remains essentially unchanged,
but our uncertainty becomes now two orders, rather than one order,
smaller than the experimental one. Our value of $D_0$ differs by  
1.8 $\sigma$ and 1.2 $\sigma$, respectively, from  the values
1.1$^{+0.3}_{-0.2}$~mK $\approx$ $95^{+25}_{-15}$~neV \cite{Grisenti:00}
and  $112^{+22}_{-16}$~neV \cite{Cencek:12,Spirko:13}
derived from a nanosieve transmission experiment  \cite{Grisenti:00}.
The value $D_0$=151.9$\pm$13.3~neV, obtained very recently~\cite{Zeller:16}
using the Coulomb explosion technique,
  agrees with our theoretical prediction  within 0.98 $\sigma$.


The interaction energies presented in this paper establish a new
accuracy benchmark for the helium dimer. This improvement was achieved
 using the  ECG approach to solve 
 the four-electron Schr\"odinger equation in the BO
approximation and by computing the post-BO
corrections using improved methodology and significantly larger basis sets.
We also computed, for the first time,
the effective $R$-dependent vibrational and rotational masses and 
the resulting nonadiabatic corrections to the properties of
the $^4$He$_2$ bound state. These calculations demonstrated that atomic masses
should be used in adiabatic calculations for weakly bound systems.
The predicted dissociation energy is in agreement with
the experimental determination via Coulomb explosion method,
confirming the reliability of this technique.
In a separate publication, we will report applications of
the computed potential and effective masses to calculate
properties of bulk helium of relevance to metrology.

\begin{acknowledgments}
  This work was supported by the NSF grant CHE-1566036 and
  the NCN grant 2014/15/B/ST4/04929.
\end{acknowledgments}

%

\end{document}